\documentclass[fleqn,usenatbib]{mnras}

\usepackage{newtxtext,newtxmath}

\usepackage[T1]{fontenc}
\DeclareRobustCommand{\VAN}[3]{#2}
\let\VANthebibliography\thebibliography
\def\thebibliography{\DeclareRobustCommand{\VAN}[3]{##3}\VANthebibliography}

\usepackage{graphicx}	
\usepackage{amsmath}	

\usepackage{caption}
\usepackage{subfigure}

\title{On the ultra-long spin period of 4U 1954$+$31 
}

\author[Mao \& Li]{
Ying-Han Mao,$^{1,2}$
Xiang-Dong Li$^{1,2}$\thanks{E-mail: lixd@nju.edu.cn}
\\
$^{1}$School of Astronomy and Space Science, Nanjing University, Nanjing 210023, P. R. China\\
$^{2}$Key Laboratory of Modern Astronomy and Astrophysics (Nanjing University),
Ministry of Education, Nanjing 210023, P. R. China}

\date{Accepted XXX. Received YYY; in original form ZZZ}

\pubyear{2024}

\begin{document}
\label{firstpage}
\pagerange{\pageref{firstpage}--\pageref{lastpage}}
\maketitle

\begin{abstract}
4U 1954$+$31 is a high-mass X-ray binary (HMXB) that contains a neutron star and an M supergiant companion. 
The neutron star has a spin period of $\sim 5.4$ hr. 
The traditional wind-accreting model requires an ultra-strong magnetic field for the neutron star to explain its extremely long spin period, which seems problematic for the neutron star with an age of a few $10^7$ years. In this work, we take into account the unsteady feature of wind accretion, which results in alternation of the direction of the wind matter's angular momentum. Accordingly, the torque exerted by the accreted wind matter varies between positive and negative from time to time, and largely cancels out over a long time. In such a scenario, neutron stars can naturally attain long spin periods without the requirement of a very strong magnetic field. This may also provide a reasonable explanation for the spin period distribution of long-period neutron stars in HMXBs.
\end{abstract}




\section{Introduction}

4U 1954$+$31 was first detected by Uhuru \citep{Forman1978}.
Based on optical observations, this system had been classified as a symbiotic X-ray binary (SyXB) that comprises a neutron star (NS) and an M4-5 III star (of mass $\sim 1.2\,{\rm M}_\odot$) at a distance of 1.7 kpc \citep{Masetti2006}.
Combining the Gaia parallax with infrared color, spectral type, abundances, and orbital properties of the optical counterpart, \cite{Hinkle2020} showed that the cool star in this system is not a
low-mass giant but an M4I supergiant of mass $9^{+6}_{-2}~{\rm M}_{\odot}$ at a distance of $\sim 3.3$ kpc.
The estimated stellar radius ($\sim 590~{R_\odot}$)  imposes the binary separation $\gtrsim {\rm5~au}$, corresponding to an orbital period longer than 3 years \citep{Hinkle2020}
Consequently, 
4U 1954$+$31 belongs to the few known red supergiant/high-mass X-ray binaries (RSG/HMXBs) in the Milky Way\footnote{The three other binaries are Scutum X$-$1, CXO 174528.79$-$290942.8, and SWIFT J0850.8$-$4219 \citep{Kaplan2007,Gottlieb2020,De2024}.}.

Apart from its unique optical properties, another remarkable feature of 4U 1954$+$31 is its 5.4 hr pulse period \citep{Corbet2006,Corbet2008}, which is the longest among all known accretion-powered pulsars. The NS
exhibited large ($\sim 7\%$) fluctuations in X-ray luminosity and alternating phases of spin-up and spin-down \citep{Marcu2011}, indicating that it is likely near its equilibrium period. 
\citet{Mattana2006} suggested that the orbital period ($P_{\rm orb}$) should be at least 400 days based on the pulse period evolution.
In the standard magnetically threaded accretion disc model \citep{Ghosh1979}, the NS's equilibrium spin period is 
\begin{equation}
    P_{\rm eq}\simeq 7.0\mu_{30}^{6/7}\dot{M}_{16}^{-3/7}\,{\rm s},
\end{equation}
where $\mu=\mu_{30}\cdot 10^{30}$ G~cm$^3$ is the NS magnetic dipole moment, and $\dot{M} =\dot{M}_{16}\cdot 10^{16}$ g~s$^{-1}$ is its mass accretion rate.
It is seen from Eq.~(1) that, to attain such a long spin period would require a magnetic field of $\sim 10^{16}$ G with $\dot{M}_{16}=0.4$ \citep{Enoto2014}.
Although magnetar-like magnetic fields have been suggested for various types of X-ray binaries \citep[][and references therein]{Xu2022,Popov2023}, they should be limited to very young systems considering the fact that the
ultra-strong fields would decay on a timescale $\lesssim 10^5-10^6$ yr \citep{Turolla2015}. 

\cite{Enoto2014} showed that the X-ray continuum of 4U 1954$+$31 is well described by a Comptonized model with the addition of a narrow 6.4 keV Fe-K$\alpha$ line during the outburst, similar to slowly rotating pulsars in HMXBs. Its high pulsed fraction ($\sim 60\%-80\%$) and the possible location in the Corbet diagram \citep{Corbet1984,Corbet1986} also favor a normally high B-field ($\sim 10^{12}-10^{13}$ G).
Taking into account its low luminosity
($\sim 0.02-2\times 10^{35}$ ergs$^{-1}$), presumably wide orbital period, and rather low wind velocity from the donor star, \cite{Enoto2014} suggested that the NS is unlikely fed by a disc but a wind accretor, and the long spin period could be explained by the subsonic settling accretion model 
\citep{Shakura2012}. The equilibrium spin period in the subsonic settling accretion regime is \citep{Shakura2012} 
\begin{equation}
    P_{\rm eq}=1.2\times 10^4\mu_{30}^{12/11}\dot{M}_{16}^{-4/11}v_8^4\left(\frac{P_{\rm orb}}{100~{\rm days}}\right)\,{\rm s},
\end{equation}
where $v_{\rm rel}=v_8\cdot 1000$ km~s$^{-1}$ is the relative velocity between
the stellar wind and the NS.
Taking $B\sim 10^{12}-10^{13}$ G, then its spin period can only be explained by a relatively fast stellar wind for late-type giants (e.g., $\sim 300$ km~s$^{-1}$). However, a terminal wind velocity of $10-30$\,km~s$^{-1}$ is expected for M4I supergiants \citep{Jura1990,Mauron2011}.
Taking $P_{\rm eq}=5.4$ hr and assuming $\dot{M}_{16}=0.4$ and $v_8 = 0.04$, one can estimate $B\sim 1.7\times 10^{16}$ G for $P_{\rm orb} = 3$ years, or $B\sim 2.5\times 10^{14}$ G for $P_{\rm orb} = 300$ years \citep{Xu2022,Bozzo2022}. 
Therefore, it is challenging to explain the spin period of 4U 1954$+$31 within the current framework of wind-accreting NSs.

In the subsonic settling accretion regime, the NS is accreting from a surrounding, quasi-spherical shell of hot material. \cite{Shakura2012} showed that, the accretion torque is partly from the falling matter and partly from plasma-magnetosphere interactions. As to the former torque component, almost all of the subsequent works implicitly assume that the specific angular momentum of the accreted wind matter is prograde, thus the transfer of angular momentum between them always spins up the NS (see Eq.~(16) below). This is based on the traditional treatment \citep[e.g.,][]{Illarionov1975} of angular momentum transfer in \cite{Bondi1944} wind accretion. However, many wind-fed X-ray pulsars in SG/HMXBs  were found to show spin-up/spin-down reversals on timescale from days to decades \citep[e.g.,][and references therein]{Bildsten1997,Malacaria2020}. Similar features have been revealed by both 2D and 3D
numerical simulations of wind accretion, indicating negligible net accretion of angular momentum onto the NSs \citep[e.g.,][]{Forman1978,Matsuda1987,Fryxell1988, Blondin2009,Blondin2012,Xu2019}. 
The implication is that the accreted angular momentum might alternate between positive and negative values and largely cancel out each other over a long time, significantly reducing the averaged accretion torque and resulting in a longer equilibrium spin period for the NS.
This effect, however, has not been explored in previous analyses of the spin evolution of wind-fed NSs.

The following of this paper is structured as follows.
The wind accretion torque model is described in Section 2. Then it is applied to model the spin evolution of 4U 1954$+$31, and investigate the  spin period distribution of X-ray pulsars in SG/HMXBs in Section 3. We summarize our results in Section 4.

\section{The wind-accretion torque model}
\subsection{The standard evolution model}
We consider a binary system consisting of an NS with mass $M_{\rm NS}$ and a main-sequence star with mass $M_2$, which is not filling its Roche lobe. For simplicity we assume that the binary orbit is circular. The NS captures the wind material from its companion at a rate
\citep{Bondi1944}
\begin{equation}
    \dot{M}=\frac{\pi R_{\rm G}^{2}\dot{M}_{2}v_{\rm rel}}{4\pi a^{2}v_{\rm w}},
	\label{eq:quadratic}
\end{equation}
where the relative velocity $v_{\rm rel}=(v_{\rm w}^{2}+v_{\rm orb}^{2})^{1/2}$, $v_{\rm w}$ and $v_{\rm orb}$ are the stellar wind velocity and the orbital velocity respectively, $a$ is the orbital separation, $R_{\rm G}={2GM_{\rm NS}}/{v_{\rm rel}^{2}}$ is the gravitational capture radius of the NS, and $\dot{M}_{2}$ is the mass loss rate of the companion star. 
We use the standard formula from \cite{Castor1975} to estimate the wind velocity
\begin{equation}
    v_{\rm w}=v_{\rm \infty}(1-{R_{2}\over{a}})^{\beta}=\alpha v_{\rm esc}(1-{R_{2}\over{a}})^{\beta},
    \label{eq:quadratic}
\end{equation}
where $v_{\rm \infty}$ is the terminal wind velocity, $v_{\rm esc}=\sqrt{2GM_{2}/R_{2}}$ is the escape velocity from the companion star. 
For the coefficient $\alpha$ and the power law index $\beta$, we adopt $\alpha=1$ and 2, and $\beta=0.8$ \cite[]{Waters1989, Karino2020}.

The magneto-rotational evolution of a wind-fed NS is mainly determined by the interaction of the NS with the surrounding wind matter, depending on the orbital separation, the NS spin period and magnetic field, and the loss rate and velocity of the companion's wind. We refer to \cite{Abolmasov2024} for a recent review.
Generally speaking, a newborn NS is usually in the \textit{ejector} phase (or \textit{phase a}) if the magnetospheric radius $R_{\rm m}$ is greater than the light cylinder radius $R_{\rm lc}$ or the gravitational capture radius $R_{\rm G}$. In this phase the surrounding material cannot penetrate into the light cylinder and be accreted onto the NS. The magnetosphere radius is expressed in terms of the Alfv\'en radius \citep[]{Lamb1973,Fabian1975}
\begin{equation}
    R_{\rm m}\simeq R_{\rm A}=\left(\frac{\mu^2}{2\dot{M}\sqrt{2GM_{\rm NS}} }\right)^{2/7},
	\label{eq:quadratic}
\end{equation} 
where $G$ is the gravitational constant, and $\mu=BR_{\rm NS}^{3}$ (where $B$ and $R_{\rm NS}$ are the magnetic dipole field and the radius of the NS respectively).
The light cylinder radius is
\begin{equation}
    R_{\rm lc}=\frac{cP_{\rm s}}{2 \pi},
	\label{eq:quadratic}
\end{equation}      
where $c$ is the speed of light and $P_{\rm s}$ is the spin period of the NS. 

In the $\textit{ejector}$ phase, the accretion matter does not interact with the NS, so the NS spins down solely through magnetic dipole radiation. The torque exerted on the NS is
\begin{equation}
    N_{\rm dip}=-\frac{16\pi^{3}\mu^{2}}{3 c^{3}P_{\rm s}^{3}}.
	\label{eq:quadratic}
\end{equation}   

As time goes on, the magnetospheric radius $R_{\rm m}$ gradually approaches and eventually penetrates into the light cylinder. The NS enters the $\textit{propeller}$ phase (or $\textit{phase b}$) if $R_{\rm m}$ is located between the light cylinder radius $R_{\rm lc}$ and the corotation radius $R_{\rm co}$, which is defined as the radial distance where the spin angular velocity of the NS is equal to the Keplerian orbital angular velocity,
\begin{equation}
    R_{\rm co}=\left(\frac{GM_{\rm NS}P_{\rm s}^{2}}{4 \pi^{2}}\right)^{1/3}.
	\label{eq:quadratic}
\end{equation} 
Because the accretion matter is forced to be corotating with the magnetosphere at super-Keplerian speed, the centrifugal force exceeds gravity. 
As a result, the matter cannot be accreted onto the NS and is either propelled away or halted at the boundary \citep{Illarionov1975}. The spin-down torque caused by the flung out of matter is \citep{Shakura1975}
\begin{equation}
    N_{\rm pr}=-\dot{M}R_{\rm m}^{\rm 2}\left(\frac{2\pi}{P_{\rm s}}\right).
	\label{eq:quadratic}
\end{equation}  

The spin-down in the $\textit{propeller}$ phase is much more rapid than in the $\textit{ejector}$ phase, resulting in growing $R_{\rm co}$. 
Once $R_{\rm m}<R_{\rm co}$, the inflowing matter is channeled onto the polar caps of the NS by the magnetic field lines and undergoes angular momentum transfer. 
In this $\textit{accretor}$ phase there are two different accretion regimes: Bondi-Hoyle (BH) accretion ($\textit{phase c}$) and subsonic settling accretion ($\textit{phase d}$) \citep{Shakura2012}. 
When the X-ray luminosity $L_{\rm X}>L_{\rm crit}=4\times10^{36}\mu_{30}^{1/4}~{\rm erg}~{\rm  s^{-1}}$ \citep{Postnov2011}, the cooling timescale of the matter is less than its free-fall timescale.
The cooled matter free falls and enters the magnetosphere supersonically.
When the X-ray luminosity is below $L_{\rm crit}$, the Compton cooling efficiency significantly decreases. 
The radial velocity of the matter becomes subsonic, forming a hot shell outside the NS. Accretion occurs through the Rayleigh-Taylor instability. 
The torque in $\textit{phase c}$ can be written as \citep{Popov1999}
\begin{equation}
    N_{\rm super}=Z\dot{M} R_{\rm G}^{\rm 2}\left(\frac{2\pi}{P_{\rm orb}}\right)-\frac{\mu^{2}}{3R_{\rm co}^{3}},
	\label{eq:quadratic}
\end{equation}  
where $Z$ is a dimensionless coefficient, derived to be 0.25 in the case of steady wind accretion \citep[]{Illarionov1975}.
In $\textit{phase d}$, the torque is expressed as \cite[]{Popov2012}
\begin{equation}
    N_{\rm sub}=A\dot{M}_{\rm {sub,16}}^{7/11}-B\dot{M}_{\rm {sub,16}}^{3/11},
	\label{eq:quadratic}
\end{equation}  
where
\begin{equation}
A=4.60\times10^{31}K_{1}\mu^{1/11}_{30}v_{8}^{-4}\left(\frac{P_{\rm orb}}{10~\rm day}\right)^{-1},
	\label{eq:quadratic}
\end{equation} 
and
\begin{equation}
B=5.49\times10^{32}K_{1}\mu^{13/11}_{30}\left(\frac{P_{\rm s}}{100~\rm s}\right)^{-1}.
	\label{eq:quadratic}
\end{equation} 
Here $K_{1}\simeq 40$ is a dimensionless numerical factor and $\dot{M}_{\rm sub}=\dot{M}_{\rm sub,16} \cdot 10^{16}$ g~s$^{-1}$. 
In $\textit{phase~d}$, accretion matter enters the magnetosphere through Rayleigh-Taylor instability, so the accretion rate is less than the capture rate of the NS by a factor \citep{Postnov2011}
\begin{equation}
{\dot{M}_{\rm sub}\over{\dot{M}}}=0.3\dot{M}_{16}^{4/11}\mu_{30}^{-1/11}.
	\label{eq:quadratic}
\end{equation} 

\begin{figure}
	\includegraphics[width=\columnwidth]{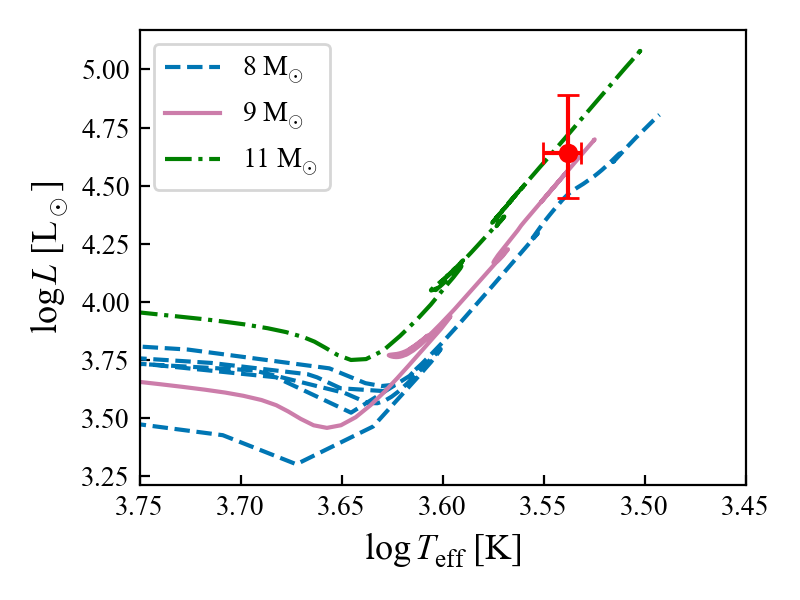}
    \caption{The evolutionary tracks on the Hertzsprung-Russell diagram of stars with different initial masses, simulated with MESA. 
    The red point with error bars represents the observed position of the optical companion star in 4U 1954$+$31.
}
    \label{fig:Figure1}
\end{figure}

\begin{figure*}
	\includegraphics[width=2\columnwidth]{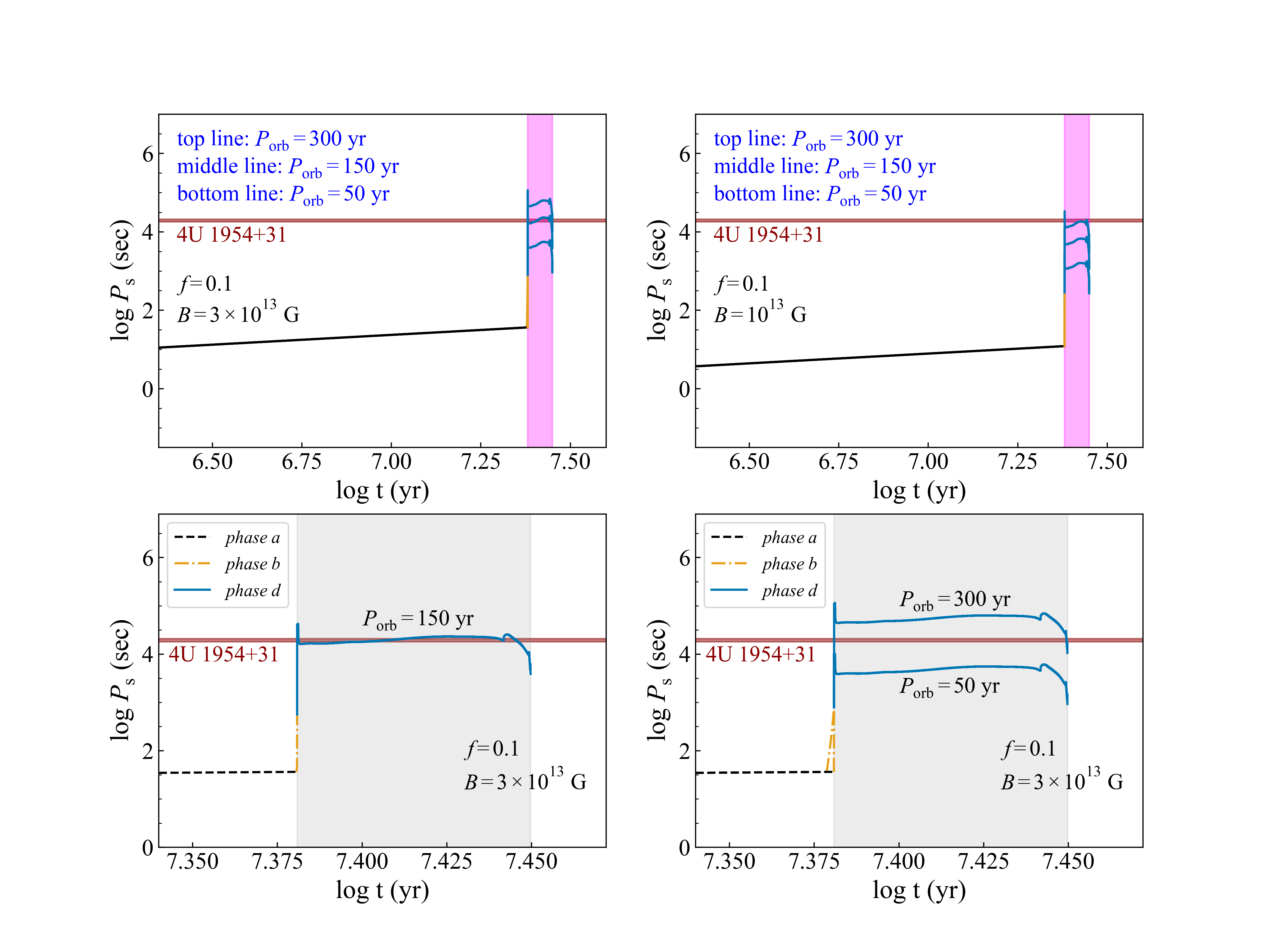}
    \caption{Top panels:
    The spin evolutionary tracks of the NS in 4U 1954$+$31 with different parameters. 
    The black, orange, and blue lines signify $\textit{phase~a}$, $\textit{b}$, and $\textit{d}$, respectively.
    The three lines from top to bottom represent the evolution with $P_{\rm orb}=300$, 150, and 50 yr, respectively, and they overlap each other before 24 Myr.
    The magenta column and the horizontal line represent the range of the possible companion's age and the measured spin period, respectively.
    The top-left and top-right panels compare the evolution tracks with different NS magnetic fields.
    Bottom panels:
    These two panels show the zoom of spin evolution within the plausible age range in the top-left panel with $B = 3 \times 10^{13}$ G. For clarity, we plot the cases of $P_{\rm orb}=150$ yr and $P_{\rm orb}=50$ and 300 yr in the left and right panels, respectively. The dashed, dot-dashed and solid lines represent $\textit{phases~a}$, $b$, and $d$, respectively.
    }
    \label{fig:Figure2}
\end{figure*}

\subsection{Unsteady wind accretion}
We see that the total torque in both supersonic and subsonic accretion cases contains spin-up and spin-down terms.
The first part of Eq.~(10) comes from angular momentum transfer between the captured matter and the NS, and the second describes the effect of magnetic braking.
In Eq.~(11), the spin-up torque originates from the interaction between the plasma shell and the NS.
Accretion matter at the inner boundary of the shell spins at a certain angular velocity $\omega_{\rm m}$ which is determined by the orbital period,  the gravitational capture radius, and the magnetospheric radius \citep{Shakura2012}
\begin{equation}
    \omega_{\rm m}=\tilde{\omega}\omega_{\rm orb}\left(\frac{R_{\rm G}}{R_{\rm m}}\right)^2
	\label{eq:quadratic}
\end{equation} 
where $\tilde{\omega}$ is a numerical factor. 
The sign and magnitude of the torque generated by the shell-magnetosphere interaction are determined by the angular velocity difference between the rotating matter and the magnetosphere
\begin{equation}
    N_{\rm inter}=Z\dot{M}R^{2}_{\rm m}(\omega_{\rm m}-\omega_{\rm s}),
	\label{eq:quadratic}
\end{equation} 
where $\omega_{\rm s}=2\pi/P_{\rm s}$ is the angular velocity of the NS.
At the magnetospheric radius, matter acquires the same angular velocity as $\omega_{\rm s}$, and while freely falling, it transfers angular momentum back to the NS.
The expression for the torque generated by this process is
\begin{equation}
    N_{\rm acc}=z\dot{M}R^{2}_{\rm m}\omega_{\rm s},
	\label{eq:quadratic}
\end{equation} 
where $z$ represents a dimensionless parameter denoting the direction of the material fall.
The total torque acting on the NS is the combination of the two torque components,
\begin{equation}
   N=N_{\rm inter}+N_{\rm acc}=Z\dot{M}R_{\rm G}^{2}\tilde{\omega}\omega_{\rm orb}-\dot{M}R_{\rm m}^{2}(Z-z)\omega_{\rm s}.
	\label{eq:quadratic}
\end{equation} 
Here we have substituted $\omega_{\rm m}$ from Eq.~(14).
After a series of derivations, this equation can be recovered to Eq.~(11), so it is seen that the first 
term on the rhs of Eq.~(18) comes from the interaction between the rotating shell and the magnetosphere, and the second term is determined by the spin of the NS and the falling matter.
Conventionally, the spin and orbital axes of accreting NSs are thought to be aligned, so as the rotational axis of the plasma shell, i.e., $\omega_{\rm m}$ is positive. 
So the first torque component consistently acts as a spin-up torque.

The standard BH accretion model assumes a point mass accreting from a supersonic flow with constant density and velocity at infinity \citep[][for a review]{Edgar2004}. For real wind accretion in HMXBs, the upstream flow may be asymmetric because of azimuthal gradient in both density and velocity. Although the structure of the RSG winds is still poorly known, the presence of clumps is strongly suggested by the observational evidence of dense sub-structures in the wind flow \citep{Mart2017,Kaminski2019} and fast X-ray flares \citep[][and discussions therein]{Ferrigno2020,Bozzo2022}. Early analytical study by \citet{Wang1981} showed that the magnitude and sign of the angular momentum captured by an NS from its massive companion star are very sensitive to the distribution of the relative flow velocity and mass density across the capture cross-section. Small fluctuation in the mass transfer process can thus lead to a reversal of the accretion torque.
Numerical simulations in 2D \citep{Matsuda1987,Fryxell1988}
revealed that the wind accretion flow is usually unstable, exhibiting a flip-flop instability. More recent 3D simulations \citep{Matsuda1991,Ruffert1994,Ruffert1995,Ruffert1997,Ruffert1999,Blondin2012,MacLeod2015,Xu2019} did not observe the flip-flop instability, but confirmed the unstable character of the flow.
Moreover, at high wind speed, accretion can still be modeled using the standard BH accretion, while at low wind speed, the upstream gradients and orbital effects are crucial, and simple BH model may not apply. 

For unstable wind accretion, the direction of the angular momentum of the captured wind material may undergo alternations from time to time, reducing the mean specific angular momentum of the flow near the accretor.
This feature has already been pointed out by \cite{Shakura2012}, ``due to inhomogeneity in the incoming flow, a non-stationary regime with an alternating sign of the captured matter angular momentum can be realized. Thus, the sign of $Z$ can be negative as well, and alternating spin-up/spin-down regimes can be observed".
The alternating sign of $\omega_{\rm m}$ implies that the averaged torque over a long timescale may be significantly smaller than the result obtained in Eq.~(16).
To account for this effect, we introduce a dimensionless factor $f$, the magnitude of which is considerable smaller than 1, to describe the net angular momentum transfer of the stellar wind, and rewrite Eqs.~(10) and (11) to be 
\begin{equation}
    \langle N_{\rm super}\rangle=fZ \dot{M} R_{\rm G}^{\rm 2} \left(\frac{2\pi }{P_{\rm orb}}\right)-\frac{\mu^{2}}{3R_{\rm co}^{3}}~~~~~~~~{\rm if}\ L_{\rm X}>L_{\rm crit},
	\label{eq:quadratic}
\end{equation} 
\begin{equation}
    \langle N_{\rm sub} \rangle=f A\dot{M}_{\rm sub,16}^{7/11}-B\dot{M}_{\rm sub,16}^{3/11}
    ~~~~~~~~~~~{\rm if}\ L_{\rm X}<L_{\rm crit}.
	\label{eq:quadratic}
\end{equation} 
The equilibrium period in $\textit{phase d}$ changes to be
\begin{equation}
    P_{\rm eq}^{\rm sub}=\left(\frac{1.2\times 10^4}{f}\right) \mu_{30}^{12/11} v_{8}^{4} \dot{M}_{\rm sub,16}^{-4/11}\left(\frac{P_{\rm orb}}{100~{\rm day}}\right)~{\rm s}.
\label{eq:quadratic}
\end{equation} 
Note that it works only when $f>0$.

The long-term spin period evolution is determined by
\begin{equation}
    \frac{2\pi I}{P_{\rm s}^{2}}\frac{{\rm d}P_{\rm s}}{{\rm d}t}=-N,
	\label{eq:quadratic}
\end{equation}  
where $I$ is the moment of inertia of the NS, taken to be $10^{45} \rm~g~cm^{2}$ in this work. 

In the following section, we adopt the modified torque model to investigate the spin evolution of 4U 1954$+$31 and other NSs in SG/HMXBs. 
Before doing that we first estimate the possible range of $f$ from both theoretical and observational perspectives. We refer to the hydrodynamic simulation work by \cite{Xu2019} (hereafter XS2019) for the former.
For finite azimuthal upstream gradients, XS2019 divided the state of the accretion flow into three regimes: stable flow (small upstream gradient), turbulent unstable flow without a disc (intermediate upstream gradient), and turbulent flow with a disc-like structure (relatively large upstream gradient). The transverse density and velocity gradient parameters, $\epsilon_\rho$ and $\epsilon_v$, are defined by
\begin{equation}
    \epsilon_\rho=R_{\rm G}\frac{\partial\ln\rho}{\partial y}, \ 
    \epsilon_v=R_{\rm G}\frac{\partial |v_x|}{\partial y}.
\end{equation} 
Here $(x, y, z)$ is the Cartesian coordinate with the $xy$ plane being the orbital plane, $+z$ aligned with the orbital angular momentum, and
$x$ aligned with the wind velocity at the NS. $\epsilon_\rho$ and $\epsilon_v$ approximately correspond to the fractional change of the density $\rho$ and velocity $v$ per $R_{\rm G}$, respectively. Most observed SG/HMXBs have $0.01<\epsilon_\rho<0.4$ and $0.002<-\epsilon_v<0.2$, placing them in the regime of a turbulent, disc-less accretion flow \citep{Xu2019}.

Let $R_{\rm in}$ represent the inner boundary of the simulation in XS 2019, and define $L_{\rm K}=(GM_{\rm NS}R_{\rm in})^{1/2}$,  we can roughly estimate the mean specific angular momentum $\langle L\rangle$ of the accreted material to be
\begin{equation}
    \frac{\langle L \rangle}{L_{\rm K}} \sim \frac{fZ\tilde{\omega} R_{\rm G}^{2}\omega_{\rm orb}}{(GM_{\rm NS}R_{\rm in})^{1/2}}
    \sim (fZ\tilde{\omega})\left(\frac{R_{\rm G}}{R_{\rm in}}\right)^{1/2}\left(\frac{R_{\rm G}\omega_{\rm orb}}{v_{\rm rel}}\right).
\end{equation} 
In the simulations of XS2019, the centroid of $\langle L \rangle/L_{\rm K}$ is less than 0.1 for turbulent unstable flow without a disc (see their Figs.~10, 12-14).  Using $R_{\rm in}/R_{\rm G}=0.01-0.04$ and the Mach number $v_{\rm rel}/R_{\rm G}^{2}\omega_{\rm orb}\sim 10$ in the simulations, we can get $f<\sim 0.1/(Z\tilde{\omega})$.

Observations of the spin frequency variations in wind-fed X-ray pulsars could also be used to constrain the possible range of $f$. For example, \cite{Liao2019} reported the spin history of the SG/HMXB OAO 1657$-$415 from 2009 to 2021. 
Over the decade-long period, this source was observed to have 19 short-term torque reversals around a spin period of $\sim 37$ s. The magnitude of the short-term spin-up/down rate $|\dot{\nu}_{\rm s}|\sim 1-8\times 10^{-12}$ Hz~s$^{-1}$, while the long-term rate $\langle\dot{\nu}_{\rm s}\rangle\sim 2\times 10^{-13}$ Hz~s$^{-1}$. If the torque state is connected with prograde/retrograde accretion\footnote{The flux of the source showed a weak correlation with the spin-up/down rate.}, this implies $f\sim 0.02-0.2$.

In the following calculations, we set $f=0.1$ as a reference value. We also discuss its impact by varying its magnitude.

\begin{figure}
	\includegraphics[width=\columnwidth]{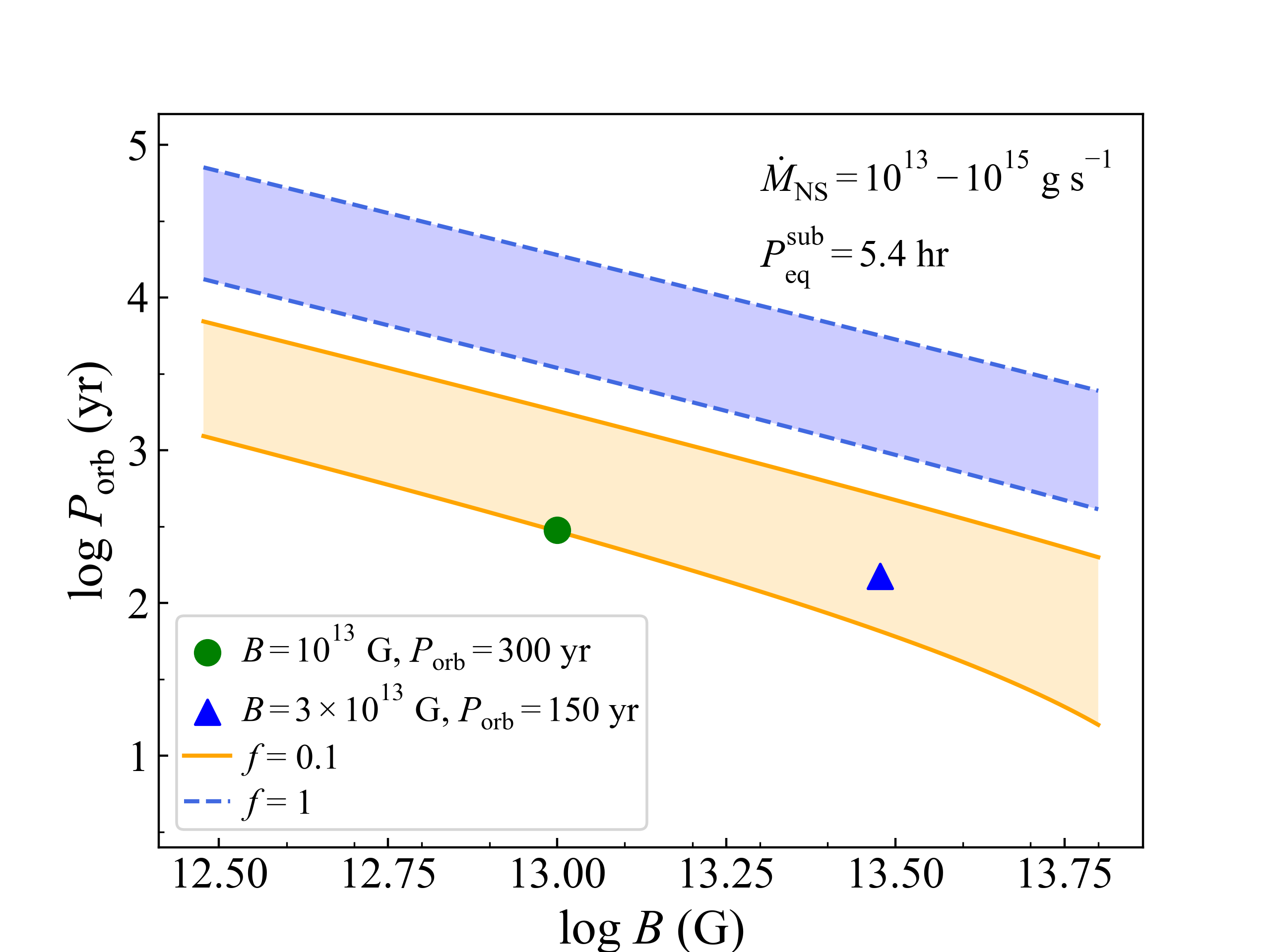}
    \caption{
    The relationship between $P_{\rm orb}$ and $B$ for 4U 1954$+$31 under the equilibrium spin condition ($P_{\rm eq}^{\rm sub} =5.4$ hr). 
    The upper and lower limits of $P_{\rm orb}$ correspond to $\dot{M}_{\rm sub} = 10^{13}$ and $10^{15} {~\rm g~s^{-1}}$. 
    The solid and dashed lines represent $f=0.1$ and 1, respectively. 
    The two symbols denote the parameters of the system matching the observed period in the top-left and top-right panels of Fig.~\ref{fig:Figure2}.
}
    \label{fig:Figure3}
\end{figure}
\begin{figure*}
	\includegraphics[width=2\columnwidth]{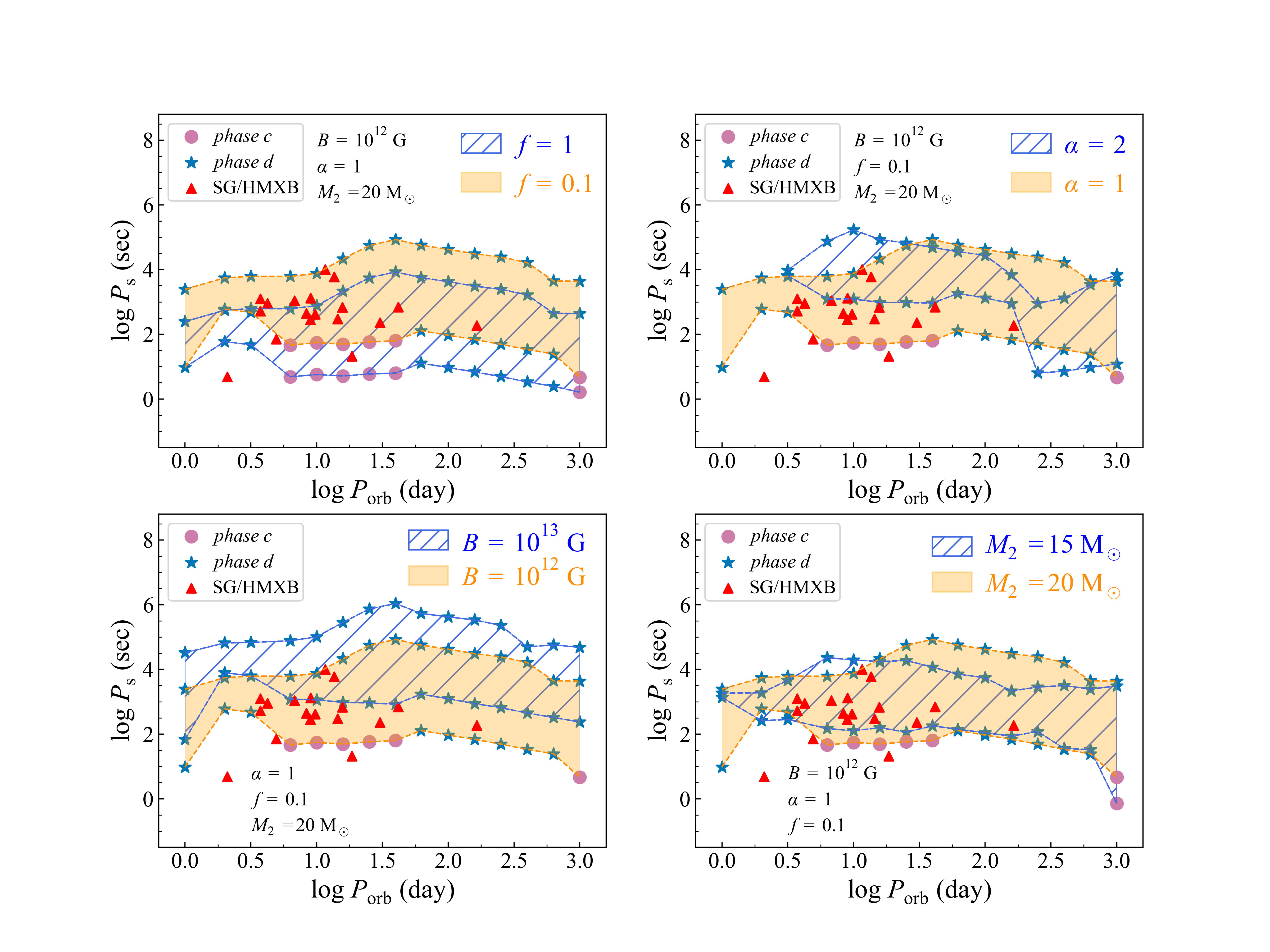}
    \caption{
    The predicted distribution of the spin periods for NSs in the $\textit{accretor}$ phase and with $L_{\rm X}>10^{34}~ \rm{erg~s^{-1}}$. 
    The pink circles and blue stars denote $\textit{phase~c}$ and $\textit{d}$, respectively.
    The red stars represent the observed X-ray pulsars in SG/HMXBs.}
    \label{fig:Figure4}
\end{figure*}

\begin{figure}
	\includegraphics[width=\columnwidth]{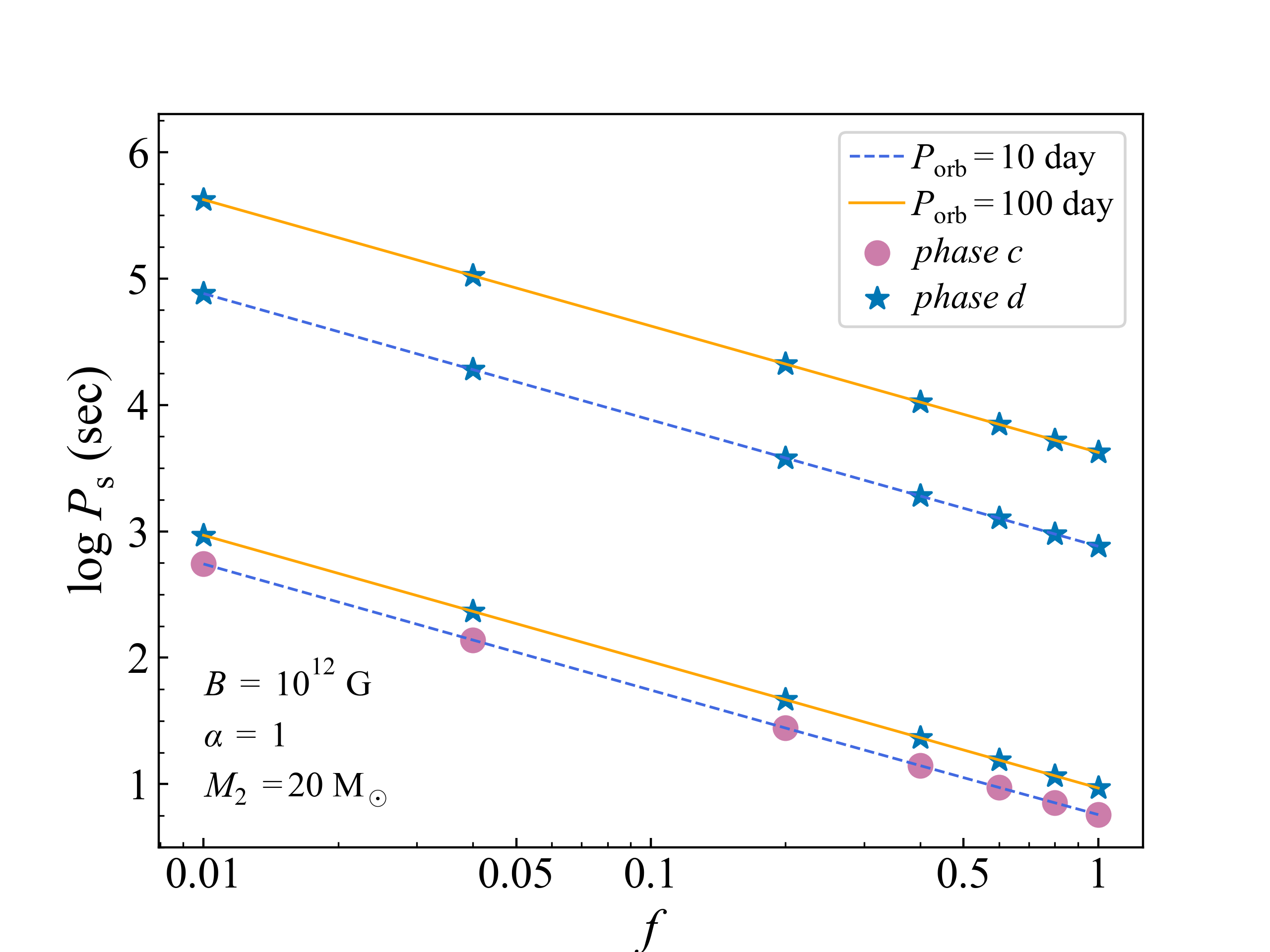}
    \caption{
    The maximum and minimum of $P_{\rm s}$ as a function of $f$.
    The results for $P_{\rm orb}$ = 10 and 100 day are represented by the blue dashed and orange solid lines, respectively.}
    \label{fig:Figure5}
\end{figure} 
\begin{figure}
	\includegraphics[width=\columnwidth]{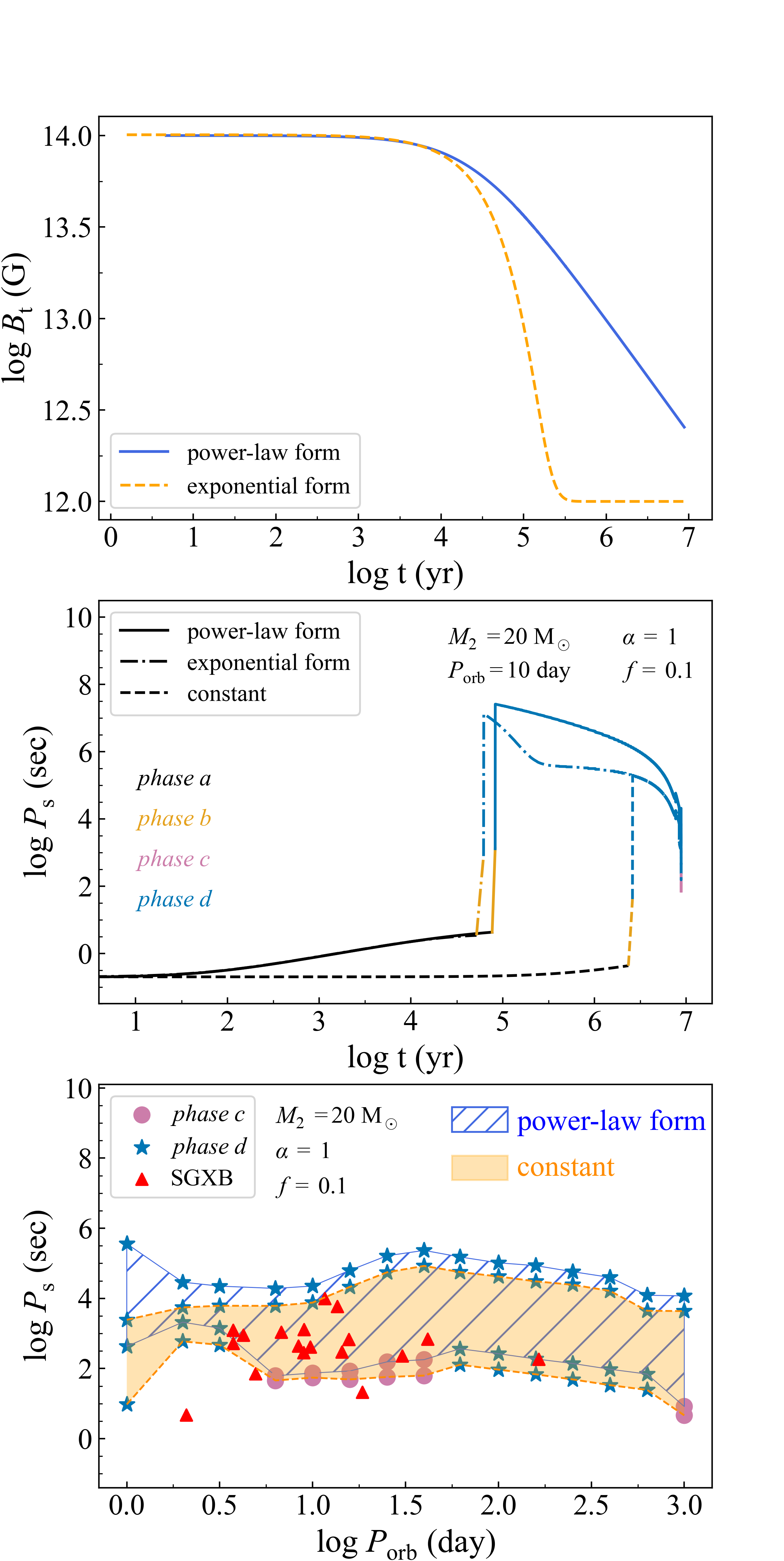}
    \caption{
    Top panel: the evolution of magnetic field strength $B$ which decays with power law and exponential form. 
    Middle panel: The solid, dashed and dot dashed lines represent the spin evolution under a constant ($10^{12}$) G magnetic field and a power-low/exponential decaying ($10^{14}$) magnetic field.
    Relevant parameters are labeled on the left side of the figure.
    Bottom panel: the spin period distributions of wind-accreting NSs with the constant and power-law magnetic fields as in the middle panel. Other parameters are the same as in Fig.~\ref{fig:Figure3}.}
    \label{fig:Figure6}
\end{figure} 

\section{Results}
 \subsection{The spin evolution of 4U 1954$+$31}

We use the stellar evolutionary code Modules for Experiments in Stellar Astrophysics (MESA) \citep{Paxton2011,Paxton2013,Paxton2015,Paxton2019} to simulate the evolution of the companion star with mass 8, 9, and 11 ${\rm M}_{\odot}$.
Its evolutionary tracks in Hertzsprung-Russell Diagram (HRD)  are shown in Fig.~\ref{fig:Figure1}.
Its current age is approximately 28 Myr for $M_2=9~{\rm M_{\odot}}$.
We employ the empirical formulas in \cite{Vink2001} and in  \cite{deJager1988} to calculate the wind loss rate for stars with $T_{\rm eff} > 1.1\times10^{4}$ K and  $T_{\rm eff} <1.1\times10^{4}$ K, respectively. The latter also applies for stars with the central hydrogen abundance < 0.01 and the central mass helium abundance by mass < $10^{-4}$. These are the default settings in the MESA code.
We use the standard wind velocity formula, Eq.~(4) (with $\alpha = 1$ and $\beta = 0.8$) for main-sequence stars.
In the RSG phase ($R > 100~{\rm R}_{\rm \odot}$), we set a typical wind velocity  $v_{\rm w} = 30~{\rm km~s^{-1}}$ \citep[]{Jura1990}.

Because the long-term evolution is insensitive to the initial value $P_{0}$ of the spin period, a fiducial value of 0.2 s is used in all calculations. 
The accreting X-ray luminosity is
\begin{equation}
    L_{\rm X}=\eta\frac{GM_{\rm NS}\dot{M}_{\rm acc}}{R_{\rm NS}},
\label{eq:quadratic}
\end{equation} 
where $\eta=0.3$ is the conversion efficiency, and the accretion rate $\dot{M}_{\rm acc}$ is calculated from the wind mass loss rate $\dot{M}_{2}$ with Eq.~(3)  (for \textit{phase~c}) and Eq.~(14) (for \textit{phase~d}).
To be compatible with the observed X-ray luminosity ($L_{\rm X}\sim 10^{33}-10^{35}~{\rm erg~s^{-1}}$), we tentatively choose the orbital period to be 50, 150 and 300 years so that $L_{\rm x}\simeq 10^{35}$, $10^{34}$, and $10^{33}~\rm {erg~s^{-1}}$ at the age of 28 Myr, respectively.
For a given orbital period, the observed X-ray variability may result from fluctuations in the mass-loss rate, eccentric orbit, random variations in the accretion rate caused by the instability of the turbulent flow \citep{Xu2019}, as well as magnetic and centrifugal gating mechanisms \citep{Bozzo2016}.

Fig.~\ref{fig:Figure2} depicts the calculated spin evolution  with $P_{\rm orb} = 50$, 150, and 300 years (corresponding to the lower, middle, and upper lines, respectively).
In both panels we set $f=0.1$, which means that $10\%$ of the instantaneous spin-up torque is on average exerted on the NS.
To better visualize the period changes after 20 Myr, the horizontal axis ($\log$\,age) of the plots starts from 6.3 in the upper panels, before which the NS is maintained in $\textit{phase~a}$ with a slow spin-down.
The magenta column depicts the possible age range of the companion star (i.e., $24-28$ Myr), approximately corresponding to the starting age of the RSG stage to the current age for a 9 ${\rm M}_{\odot}$ star.
The thick horizontal line shows the the observed spin period of 5.4 hr. 
The lower panels show the zoom of the spin evolution within the plausible age range.
 
The left panel of Fig.~\ref{fig:Figure2} shows the spin evolution for a NS with $B = 3 \times 10^{13}$ G. The different colors in the evolutionary tracks represent different phases of the NS spin evolution - the black, orange, and blue lines correspond to $\textit{phases~a}$, $b$, and $\textit{d}$, respectively.
The NS is initially born in $\textit{phase a}$, slowly spins down until the companion star becomes a RSG and the spin period reaches the critical period determined by $R_{\rm m} = R_{\rm lc}$.
After that it enters $\textit{phase~b}$, and the spin-down rate significantly increases, causing rapid growth in the spin period.
Once the period increases to the value where $R_{\rm m} = R_{\rm co}$, and $L_{\rm x}<L_{\rm crit}$, the NS enters $\textit{phase d}$ and spins down until the spin period reaches $P^{\rm sub}_{\rm eq}$.
The spin period then fluctuates around $P^{\rm sub}_{\rm eq}$ for a few Myr, and finally decreases due to the increase in the mass loss rate.
For the parameter settings in the left panel, the observed pulse period can be reached within the age range in the cases of $P_{\rm orb}=150$ and 300 yr, while
in the case of $P_{\rm orb}=50$ yr, the spin period is always lower than the observed value due to the higher mass accretion rate.

The right panel of Fig.~\ref{fig:Figure2} shows the spin evolution for a NS with $B = 10^{13}$ G.
A smaller magnetic field leads to a decrease in both the accretion torque and the equilibrium period. Thus, the curves shift downward relative to those in the left panel.
The NS can reach the observed period when $P_{\rm orb}=300$ yr.

From the spin evolution curves, we can see that once the NS enters $\textit{phase~b}$, its spin period rapidly increases and then reaches the equilibrium period in a short time.
Assume that the current spin of 4U 1954$+$31 is near the equilibrium period, we show the required $P_{\rm orb}$ as a function of $B$ for $f=0.1$ and 1 in Fig.~\ref{fig:Figure3}.
The upper and lower boundaries of the shaded region in each case correspond to the accretion rates of $10^{13}$ and $10^{15} {\rm ~g~s^{-1}}$, respectively.
The green circle and blue triangle indicate the positions for the two specific spin evolution cases in Fig.~\ref{fig:Figure2} which match the observed period.
As shown in Fig.~\ref{fig:Figure3}, an equilibrium period of 5.4 hours would still require an orbital period of several decades or over a hundred years even with $f=0.1$, unless the mass loss rate of the donor is significantly lower than our adopted value.
In recent years, Gaia observed quite a few binary systems with orbital periods longer than one year \citep[e.g.,][]{El-Badry2023,El-Badry2024}. 
Due to the sensitivity of high-precision astrometry to systems with wide orbital periods, using Gaia to constrain the long orbital period of 4U 1954$+$31 might be a feasible method to constrain the binary parameters in the future.

\subsection{The spin period distribution in SG/HMXBs}

Fig.~\ref{fig:Figure4} shows the spin period distribution of wind-fed NSs in $P_{\rm s}-P_{\rm orb}$ diagram. 
The evolution starts at the birth of the NS and stops when the companion star's radius $R_{2}$ equals to its Roche lobe radius $r_{\rm L}$.
Similar to the evolution trajectory of 4U 1954$+$31 shown above, the spin periods increase in the early stage and decrease after reaching their peak values due to the increasing mass loss rate of the companion stars.
For each specific orbital period, we record the spin periods when the NS is in the $\textit{accretor}$ phase with $L_{\rm x}>10^{34}~\rm {erg~s^{-1}}$, and depict their maximum and minimum values using different symbols to denote the evolutionary phases of the NSs - the
pink circles and blue stars represent $\textit{phases~c}$ and $\textit{d}$, respectively.
The blue-hatched and orange-shaded areas delineate the spin period range that the NSs in SG/HMXBs can attain in the $P_{\rm s}-P_{\rm orb}$ plane with $f=1$ and 0.1, respectively.
For comparison, the red triangles in the figure mark the observations of $P_{\rm s}$ and $P_{\rm orb}$ for known wind-fed X-ray pulsars with SG companions \citep[data are taken from][]{Neumann2023}.

In the reference model, we set 
$M_{2} = 20~{\rm M}_{\odot}$, $B = 10^{12}$ G, $f = 0.1$, and $\alpha = 1$, and show the results with the orange-shaded region. Although we focus on SG/HMXBs, the properties of the NS and the companion stars, as well as the wind structure may considerably change from source to source. We thus use the blue-hatched regions in the four panels in Fig.~\ref{fig:Figure4} to illustrate the impact of $f$, $\alpha$, $B$, and $M_{2}$ on the spin period range.

The top-left panel compares the spin period distributions with $f = 0.1$ and 1.
Note that the NSs always reach their minimum of $P_{\rm s}$ near the end of their evolution when the companion star starts to overflow its Roche lobe. Accordingly, the evolutionary state of the companion star at this time is largely dependent on the orbital period. For binaries with $P_{\rm orb}\lesssim 2$ days, the companion star fills its Roche lobe early, so its winds are still relatively weak at that time, leaving the NS in $\textit{phase~d}$. For binaries with 6 days $\lesssim P_{\rm orb}\lesssim 40$ days, the companion star is more evolved and the NS is able to enter $\textit{phase~c}$ because of the stronger winds. With increasing $P_{\rm orb}$, the accretion rate decreases, and the NS can only evolve to $\textit{phase~d}$. But for very long $P_{\rm orb}$, the wind loss rate becomes high enough to allow efficient accretion in $\textit{phase~c}$. On the other hand, the maximum of $P_{\rm s}$ represents the equilibrium period in $\textit{phase~d}$ in all cases.
It is seen that, given the orbital period, the spin periods in the case of $f=0.1$ are roughly an order of magnitude longer than in the case of $f=1$.
For the parameters we set, the results with $f=0.1$ better match the observations compared with $f=1$, especially for the long-spin period NSs.

For different types of early-type O/B donor stars, the escape velocity of the stellar wind is proportional to the surface temperature \citep{Kudritzki2000}. 
Hotter O supergiants have faster wind and smaller gradients compared to B stars \citep{Mellah2017}.
The top-right panel shows the influence of the wind parameter $\alpha$ on the $P_{\rm s}-P_{\rm orb}$ distribution. Because the accretion rate decreases significantly with increasing wind velocity ($\varpropto v_{\rm w}^{-4}$), a larger $\alpha$ means a lower accretion rate. Roche lobe overflow occurs early in narrow binaries, before which the mass loss rate is not high enough to power the X-ray luminosity above $10^{34}~{\rm erg}~{\rm s^{-1}}$, so there are no systems with $P_{\rm orb}<3$ day when $\alpha=2$.
For the same reason, no systems can enter $\textit{phase~c}$ during the evolution.
A larger $\alpha$ results in an increase in the equilibrium period. So the minimum of $P_{\rm s}$ with $\alpha = 2$ is overall longer than that with $\alpha = 1$. 

When the NS magnetic field is increased to $B=10^{13}$ G, the equilibrium period becomes longer, as illustrated in the bottom-left panel. 
The enhanced magnetic field also causes the accretion rate of NS to be lower in $\textit{phase~d}$ and $L_{\rm crit}$ to be higher. Thus, the NS is not able to enter $\textit{phase~c}$.

In the bottom-right panel, we consider the results with $M_2=15~{\rm M}_{\odot}$. Compared with the case of $M_2=20~{\rm M}_{\odot}$, this is equivalent to reducing the mass loss rate from the companion star.
Thus, the NSs remain in  $\textit{phase~d}$, and 
the blue-hatched region can also cover most of the observed SG/HMXBs.

To demonstrate the impact of $f$ in more detail, we select two specific systems in the upper-left panel of Fig.~\ref{fig:Figure4} with $P_{\rm orb}=10$ and 100 days, and calculate the maximum and minimum of $P_{\rm s}$ as a function of $f$. 
The results are shown in Fig.~\ref{fig:Figure5}.
The maximum and minimum values of $P_{\rm s}$ are inversely correlated with $f$, consistent with the expectation from Eqs.~(19) and (20).
This also indicates that the maximum $P_{\rm s}$ are close to the equilibrium period.  
Since the equilibrium period is often used to estimate the magnetic field strength of NSs, using the equilibrium period formula with $f$=1 might lead to an overestimation of the NS magnetic field strength.

Finally, we consider the possibility that the NSs may possess ultra-strong magnetic fields at birth, which decay  due to Hall drift and Ohmic diffusion \citep[e.g.,][]{Jones1988,Turolla2015,Xu2022,Popov2023}.
Compared with isolated NSs, accreting NSs might experience more rapid magnetic field decay due to thermo-magnetic effects \citep{Geppert1994}. 
We set an initial magnetic field of $B_{0}=10^{14}~\rm{G}$, and let it decay in a power-law form or a more rapid exponential form \citep{Colpi2000}
\begin{equation}
    B(t)=B_{0}\left (1+\gamma t/\tau \right)^{-1/\gamma},
	\label{eq:quadratic}
\end{equation}
\begin{equation}
    B(t)=B_{0}e^{-t/\tau}+B_{\rm m},
	\label{eq:quadratic}
\end{equation}
where $\gamma=1.6$, $\tau=\tau_{\rm d}/(B_{0}/10^{15}~\rm G)^{\gamma}$, $\tau_{\rm d}=10^{3}~\rm yr$, and $B_{\rm m} = 10^{12}$ G is the minimal magnetic field   \citep{Dall’Osso2012}.
The evolution of the magnetic field is shown in the top panel of Fig.~\ref{fig:Figure6}.
In the middle panel, the solid, dashed, and dot-dashed lines represent the spin evolution with a normal, constant magnetic field $B=10^{12}$ G and an initial magnetic field $B_0=10^{14}$ G decaying with the power-law and exponential form, respectively. 
A stronger initial magnetic field causes the NS to enter $\textit{phase~b}$ at around $\lesssim 10^5$ years, significantly shorter than the time (a few $10^6$ years) with $B=10^{12}$ G.
The two field decay models cause some differences in the early stages of the NS evolution (e.g., the starting time of  $\textit{phase~b}$, the spin period in the first Myr), but the duration is rather short, and the evolutions eventually align with that in the constant field model. The main reason is that the evolutionary timescale of the donor star in HMXBs is significantly longer than the field decay timescale for magnetars.
The bottom panel compares the distributions of accreting NSs in HMXBs in the $P_{\rm s}-P_{\rm orb}$ diagram. 
The magnetic field decay profiles have little effect on the  $P_{\rm s}-P_{\rm orb}$ distribution; this relationship is primarily sensitive to the final magnetic field at the late stages of evolution.
For the exponential field decay, the magnetic fields quickly reach $B_{\rm m}=10^{12}$ G, so the $P_{\rm s}-P_{\rm orb}$ distribution completely overlaps that with a constant magnetic field.

\section{Conclusions}
In this study, we investigate the spin evolution of the accreting NS in 4U 1954$+$31.
4U 1954$+$31 is a binary system with an RSG companion star that has not filled its Roche lobe. 
The NS in this system is expected to be wind-fed and near the spin equilibrium state. Based on its observed luminosity and spin period, the NS is most likely in the subsonic settling accretion state.

The NS has an ultra-long spin period of $5.4$ hours which is challenging to explain with the traditional wind-accretion torque models. 
In these models, it is generally assumed that the specific angular momentum of the accreting wind matter is always prograde, resulting in spin-up of the NS. However, both observations and numerical simulation indicate that this may not be a valid assumption in the case of unsteady wind accretion, which experiences alternations in the angular momentum at times.
We introduce a factor $f$ to average the long-term torque of the wind material for both supersonic accretion and subsonic setting accretion regimes. 
A small value of $f$ implies that the spin-up torque acting on the NS is much smaller than in solely prograde accretion. 
Based on 3D numerical simulations of BH accretion and the observed spin variations in X-ray pulsars in SG/HMXBs, we infer the plausible magnitude of $f$ and show that it may be considerably smaller than unity.
We investigate the spin history of the NS in 4U 1954$+$31 with the modified torque model. 
Assuming $f = 0.1$, we find that 4U 1954$+$31 can attain its current spin period with a normal magnetic field $\sim 10^{13}$ G. Thus, accreting NSs with ultra-long spin periods do not necessarily require ultra-strong magnetic fields. We also show that, the distribution of observed wind-accreting X-ray
pulsars in SG/HMXBs in the $P_{\rm s}-P_{\rm orb}$ plane can be reasonably reproduced with typical physical parameters.

\section*{Acknowledgments}
\quad We are grateful to an anonymous referee for helpful comments. This work was supported by the National Key Research and Development Program of China (2021YFA0718500), the Natural Science Foundation of China under grant No. 12041301, 12121003 and 12203051.

\section*{Data Availability}
\quad All data underlying this article will be shared on reasonable request to the corresponding author.



\bibliographystyle{mnras}
\bibliography{cite} 






\bsp	
\label{lastpage}
\end{document}